\documentclass{elsart}
\usepackage{natbib}
\usepackage{psfig}
\input{epsf}
\begin{document}
\runauthor{Li and Hertz}
\begin{frontmatter}
\title{Odor recognition and segmentation by coupled olfactory bulb and
cortical networks}
\author[MIT]{Zhaoping Li\thanksref{Someone}}
\author[Nordita]{John Hertz}

\address[MIT]{CBCL, MIT, Cambridge MA 02139 USA}
\address[Nordita]{Nordita, Blegdamsvej 17, DK-2100 Copenhagen {\O}, Denmark}
\thanks[Someone]{Pesent address: Gatsby Comput. Neurosci. Unit, University College, London, UK}
\begin{abstract}
We present a model of a coupled system of the olfactory bulb and
cortex.  Odor inputs to the epithelium are transformed to 
oscillatory bulbar activities. The cortex recognizes the 
odor by resonating to the bulbar oscillating pattern
when the amplitude and phase patterns from the bulb match 
an odor memory stored in the intracortical synapses.  
We assume a cortical structure
which transforms the odor information in the
oscillatory pattern to a slow DC feedback signal to the bulb.
This feedback suppresses the bulbar 
response to the pre-existing odor, allowing 
subsequent odor objects to be segmented out for recognition.
\end{abstract}
\end{frontmatter}

\section{Introduction}
\typeout{SET RUN AUTHOR to \@runauthor}

There is a great deal of current interest in how neural systems,
both artificial and natural, can use top-down feedback to modulate input 
processing. Here we propose a minimal model for an olfactory 
system in which feedback enables it to perform an essential 
task -- olfactory segmentation. Most olfactory systems need to detect,
recognize, and  segment odor objects.
Segmentation is necessary because different odors give overlapping 
activity patterns on odor receptor neurons, of which there are hundreds of 
types \cite{BuckAxel}, and each has a broad spectrum of response to 
different odor molecules \cite{Shepherd90}. 
Different odor objects seldom enter the environment in
the same sniff cycle, but they often stay 
together in the environment afterwards.  
Humans usually can not identify the individual odor objects in 
mixtures \cite{Moncrieff}, although they easily perceive an incoming
odor superposed on pre-existing ones.
Our model performs odor segmentation temporally: First one odor is 
detected and encoded by the olfactory bulb and recognized by the associative 
memory circuits of the olfactory cortex.  Then the cortex gives an 
odor-specific feedback to the bulb to inhibit the response or adapt to this odor, 
so that a superposed second odor arriving later can be detected and recognized 
with undiminished sensitivity while the sensitivity to the pre-existing odor 
is reduced, as observed 
psychophysically \cite{Moncrieff}.
The stimulus-specific feedback makes odor adaptation an 
intelligent computational strategy, unlike simple fatigue, which is not 
sufficient for odor segmentation. 
Our model displays the
oscillatory neural activities in the bulb and cortex as observed
physiologically \cite{FS}. 
Furthermore, odor cross-adaptation --- the suppression and distortion of 
odor perception immediately after an exposure to another odor --- as observed 
psychophysically \cite{Moncrieff}, is a consequence of this model.

\section{The Model}

Our model (Fig.\ 1) describes the essential elements of primary olfactory 
neural circuitry: the olfactory bulb, the olfactory cortex, 
and feedforward and feedback coupling between them.  
The formal neurons in 
our system model the collective activity of local populations of real 
neurons.  The synaptic architecture is consistent with the known physiology 
and anatomy of the olfactory system in most mammalian species \cite{Shepherd7990}.

\begin{figure}[h]
\begin{picture}(200, 200)
\put(-155, -730)
{\epsfxsize=700pt \epsfbox{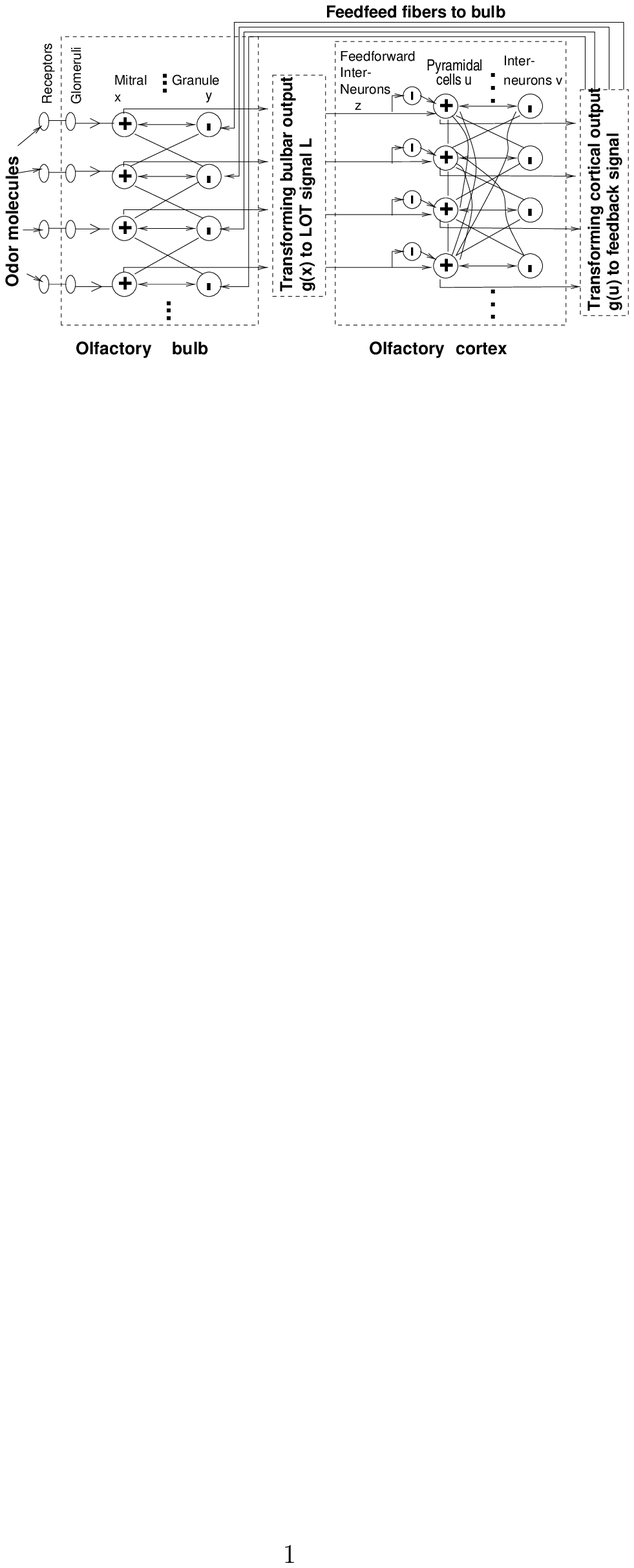}}
\end{picture}
\caption{ \label{fig:olfactorysystem}
The olfactory system in the model.} 
 
\end{figure}

Our bulb model contains interacting excitatory mitral and 
inhibitory granule cells, with membrane potentials $x_i$ and $y_i$ respectively, and
firing rates $g_x(x_i)$ and $g_y(y_i)$ respectively 
(see \cite{LH} and \cite{Li90} for details).
The odor input $I_i$ drives the dynamics 
\\
\\
$
\dot x_i = - \alpha x_i -\sum_j H^0_{ij}g_y(y_j)+ I_i   ~~~~~~~~
\dot y_i = - \alpha y_j  +\sum_j W^0_{ij}g_x(x_j) + I^c_i, 
$
\\
\\
where $-\alpha x_i$ and $-\alpha y_i$  model the decays to resting 
potentials,  ${\sf H}^0_{ij}>0$ and ${\sf W}^0_{ij}>0$  the 
synaptic connections from the granule to mitral cells and vice versa, and
vector ${\bf I}^c$ (components $I^c_i$) the feedback signal from 
the cortex to the granule cells.  
Slowly varying input $\bf I$ and ${\bf I}^c$ 
adiabatically determine the fixed or equilibrium point 
${\bf \bar x}$ and ${\bf \bar y}$ of the equations.  
Neural activities oscillates around this equilibrium as
${\bf x} = {\bf \bar x}
+\sum_k c_k {\bf X}_k e^{-\alpha t \pm {\rm i}(\sqrt \lambda _k t
+ \phi_k) }$,  where ${\bf X}_k$ is
an eigenvector of ${\sf A} = {\sf HW}$ with eigenvalue $\lambda _k$, and
$H_{ij} = H^0_{ij}g'_y(\bar y_j)$ and $W_{ij} = W^0_{ij}g'_x(\bar x_j)$.
Spontaneous oscillation occurs if ${\rm Re}(-\alpha \pm i\sqrt \lambda_k ) 
>0$; then the fastest-growing mode, 
call it ${\bf X}_1$, dominates the output and the entire bulb 
oscillates with a single frequency $\omega _1 \equiv {\rm Re} (\sqrt {\lambda_1 })$,
and the oscillation amplitudes and phases is approximately the complex vector $X_1$.
Thus, the bulb encodes the input via the steps:
(1) the input $\bf I$ determines $({\bf \bar x}, {\bf \bar y})$,
which in turn (2) determines the matrix ${\sf A}$, which then (3) determines 
whether the bulb will give spontanous oscillatory outputs and, if it does, 
the oscillation pattern ${\bf X}_1$ and frequency 
$\omega_1$.

The mitral cell outputs $g_x(x_i)$ are transformed to an effective
input $I^b_i$ to the excitatory (pyramidal) cells of the cortex by
(1) a convergent-divergent bulbar-cortex connection matrix and
(2) an effective high-pass filtering via feedforward interneurons in the
cortex. Our cortical model is structurally similar to that of the
bulb.  We focus only on the upper layer pyramidal cells and
feedback interneurons: \\ \\
$\dot u_i = -  \alpha u_i -\beta^0 g_v(v_i)
                        + \sum_j J^0_{ij} g_u(u_j)
                        + I^b_i, ~~~~
\dot v_i = - \alpha v_i + \gamma^0 g_u(u_i)
                        +\sum_j \tilde W^0_{ij} g_u(u_j)$, \\
\\
where $\bf u$, $\bf v$, and ${\sf \tilde W}^0$
correspond to $\bf x$, $\bf y$, and  ${\sf W}^0$ for the bulb.
${\sf J}^0$ is global excitatory-to-excitatory connections,
$\beta^0$ and $\gamma^0$ are local synaptic couplings.

Carrying out the same kind of linearization around the fixed point
$({\bf \bar u}, {\bf \bar v})$ as in the bulb, we obtain a 
system of driven coupled oscillators. With appropriate 
cell nonlinearities and overall scale of the synaptic connections, the
system does not oscillate spontaneously, nor does it 
respond much to random or irrelevant inputs.
However, the cortex
will resonate vigorously when the driving oscillatory force  
${\bf I}^{b}$ matches one of 
intrinsic oscillatory modes ${\vec \xi}^\mu$ in frequency and 
patterns amplitudes and phases.
These intrinsic modes ${\vec \xi}^\mu$ for $\mu = 1, 2, ...P$,
are memory items in an associative memory system 
\cite{Haberly85, WB92, AGL}, and can be
stored in the synapses 
${\sf J}^0$ and ${\sf \tilde W}^0$ in a generalized Hebb-Hopfield fashion
 
\begin{center}
$J_{ij}^0  - {{\rm i}\over \omega } ( 
	\beta \tilde W_{ij}^0 - \alpha  J_{ij}^0)
        = J \sum_\mu \xi^\mu_i \xi^{\mu*}_j/g'_u({\bar u}_j). $
\end{center}
 
\begin{figure}[h]
\begin{picture}(200, 250)
\put(-105, -590)
{\epsfxsize=630pt \epsfbox{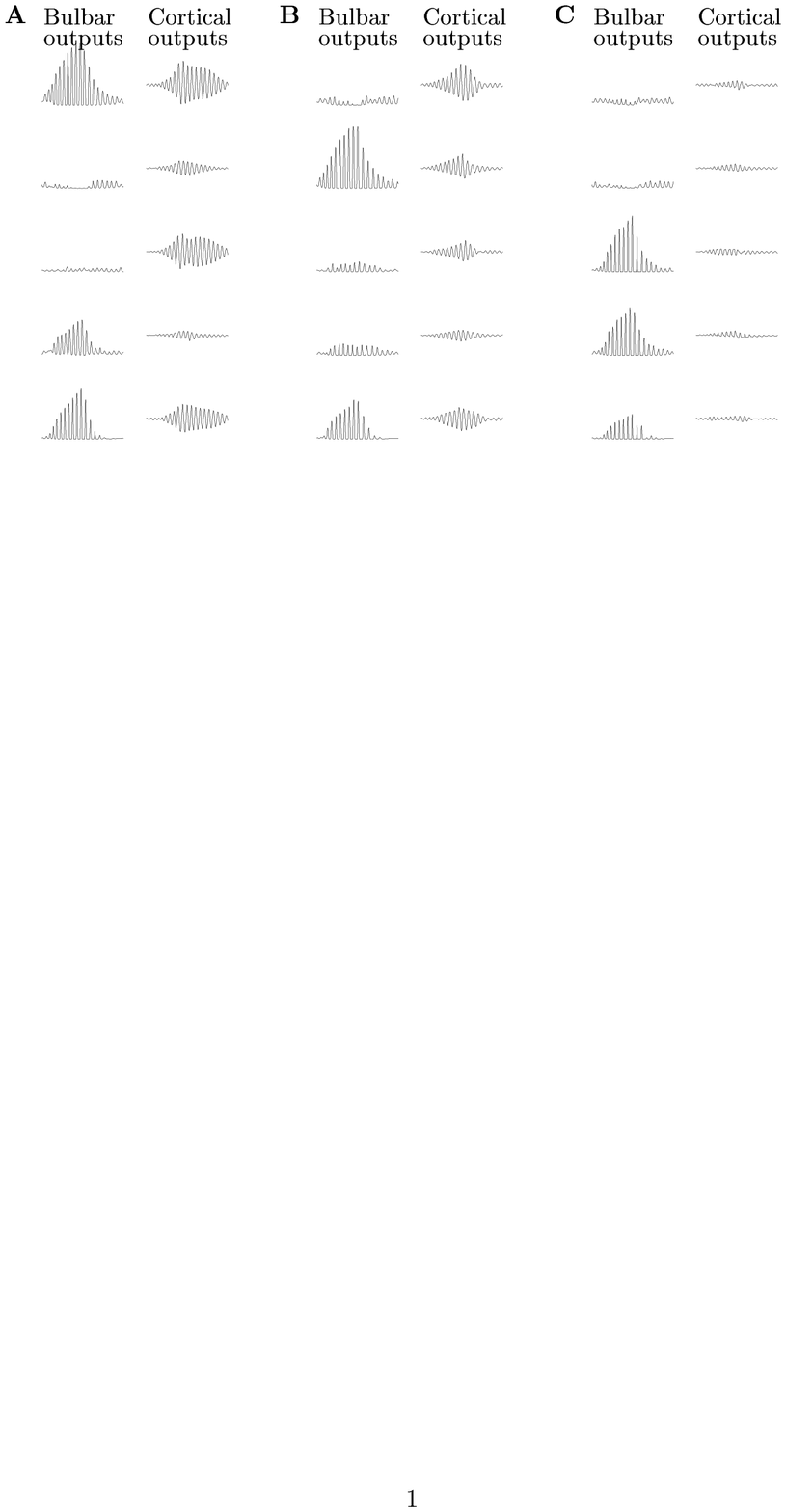}}
\end{picture}
\caption{ \label{fig:patterns} {\bf A, B, C}: bulbar and cortical oscillation
patterns for odors A, B (stored) and C (not stored)
for 5 of the 50 mitral and cortical excitatory neurons
in the model. The cortex-to-bulb feedback is turned off in the simulation
for simplicity. }
\end{figure}

Fig. \ref{fig:patterns} shows that 3 odors A, B, and C
all evoke bulbar oscillatory responses. However only odor A and B
are stored in the in the cortical synapses; hence the cortical
oscillatory response to odor C is almost nonexistent.

It was shown in \cite{Li90} that a suitable DC feedback signal to suppress 
the odor-specific activity in the bulb is 
$d{\bf I}^c = {\sf H}^{-1}\alpha d{\bf I}$. 
Somehow, this feedback should be constructed from the cortical outputs that 
contains the odor information.  We do not know how this is done in cortical
circuitry, so we treat this part of the problem phenomenologically. First,
we transform the AC signal in the pyramidal cell output $g_u(u_i)$ to a
slow DC like signal by thresholding  $g_u(u_i)$ and then
passing it through two successive very slow leaky integrators.
One can then easily construct a synaptic connection matrix to transform 
this signal to the desired feedback signal for the odor input that evoked
the cortical output $g_u(u)$ in the past sniffs. 

\begin{figure}[h]
\begin{picture}(200, 250)
\put(-105, -590)
{\epsfxsize=630pt \epsfbox{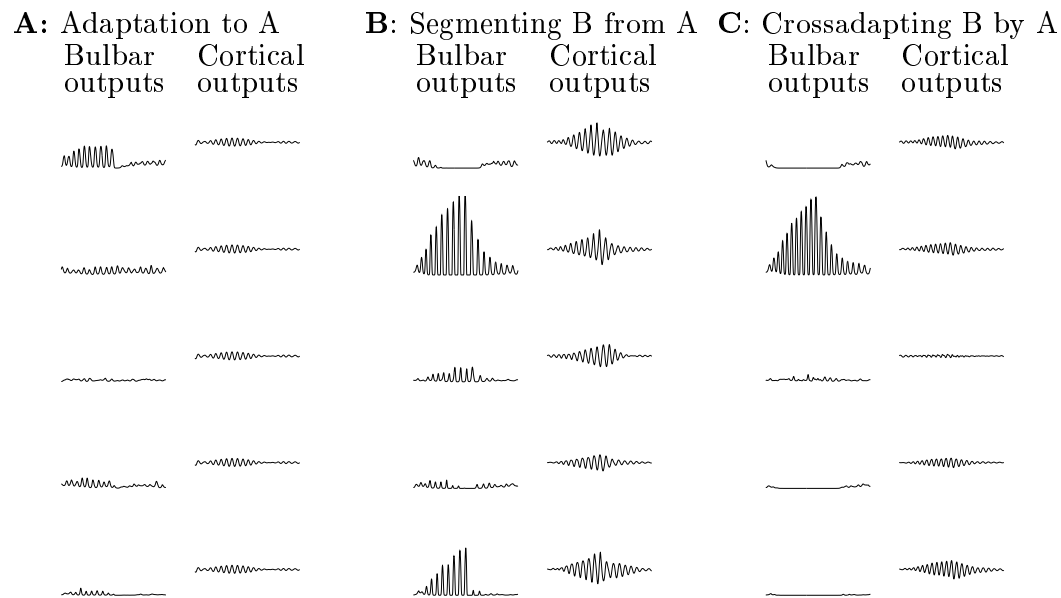}}
\end{picture}
\caption{ \label{fig:feedbackA3B1} When the feedback is turned on,
bulbar and cortical oscillation patterns for three successive
sniffs. Only odor A is present in the first two sniffs, odor B 
is present at the 3rd sniff. {\bf A}: response
to odor A in 2nd sniff, note the reduction in response levels.
{\bf B}: response to odor B superposed on odor A
in 3rd sniff, resembling that to odor B alone.
{\bf C}: response to odor B at the 3rd sniff when
odor A has been withdrawn. Note the distortion in response.
} 
\end{figure}

Feedback signal slowly builds up and the adaptation
to odor A  becomes effective at the second
sniff (Fig. \ref{fig:feedbackA3B1}A), and the system responds to 
odor A+B at the third sniff in a way as if only odor B were present
(Fig. \ref{fig:feedbackA3B1}B), 
achieving odor adaptation and segmentation consistent with human
behavior.
Quantitative analysis confirms that 
the response to the segmented odor B in the third sniff
is about 98\% similar to that of response to odor B alone.
Simulations show that odor adaptation eventually achieves 
an equilibrium level when insignificant residual responses to background
odors maintain a steady feedback signal.
A consequence of the model is olfactory cross-adaptation, when
the background odor A is suddenly removed and odor B is presented.
The feedback signal or background adaptation to odor A persists for a while and 
significantly distorts (and suppresses) the response to, and thus 
the percept of, odor B (Fig. (\ref {fig:feedbackA3B1}C)), 
as observed psychophysically \cite{Moncrieff}.

\section{Discussion}
  
We have augmented the bulb model developed in earlier work by one of
us \cite{LH,Li90} with a model of the pyriform cortex 
and with feedforward and feedback connections between it
and the bulb.  It is a minimal computational model for how
an olfactory system can detect, recognize and segment odors.
As far as we know, this is the simplest system consistent
with anatomical knowledge that can perform these three tasks,
all of which are fundamental for olfaction. 
Our model does not
deal with other computational tasks, such as hierachical catagorization
of odors \cite{AGL}.

The resonant associative memory
recognition mechanism and the slow feedback to the granule
(inhibitory) neurons of the bulb are essential parts of our model,
but many of the details of the present treatment are not.
For example, the slow feedback signal could be implemented by 
many other mechanisms, but it must be slow.  These essential features
are necessary in order that the model be
consistent with the observed phenomenology of the olfactory system.

\end{document}